# Dynamics and control of gold-encapped gallium arsenide nanowires imaged by 4D electron microscopy

Bin Chen,[1,*] Xuewen Fu,[1] Jau Tang,[1,*] Mykhaylo Lysevych,[2] Hark Hoe Tan,[3] Chennupati Jagadish,[3] Ahmed H. Zewail[1,†]

[1]Physical Biology Center for Ultrafast Science and Technology, Arthur Amos Noyes Laboratory of Chemical Physics, California Institute of Technology, Pasadena, CA 91125, USA

[2]Australian National Fabrication Facility, Research School of Physics and Engineering, The Australian National University, Canberra, ACT 2601, Australia

[3]Department of Electronic Materials Engineering, Research School of Physics and Engineering, The Australian National University, Canberra, ACT 2601, Australia

*Corresponding author. Email: bchen5@caltech.edu, jautang@caltech.edu

[†]Deceased

**Abstract:** Eutectic related reaction is a special chemical/physical reaction involving multiple phases, solid and liquid. Visualization of phase reaction of composite nanomaterials with high spatial and temporal resolution provides a key understanding of alloy growth with important industrial applications. However, it has been a rather challenging task. Here we report the direct imaging and control of the phase reaction dynamics of a single, as-grown free-standing gallium arsenide nanowire encapped with a gold nanoparticle, free from environmental confinement or disturbance, using four-dimensional electron microscopy. The non-destructive preparation of as-grown free-standing nanowires without supporting films allows us to study their anisotropic properties in their native environment with better statistical character. A laser heating pulse initiates the eutectic related reaction at a temperature much lower than the melting points of the composite materials, followed by a precisely time-delayed electron pulse to visualize the irreversible transient states of nucleation, growth and solidification of the complex. Combined with theoretical modeling, useful thermodynamic parameters of the newly formed alloy phases and their crystal structures could be determined. This technique of dynamical control and 4D imaging of phase reaction processes on the nanometer-ultrafast time scale open new venues for engineering various reactions in a wide variety of other systems.

**Significance Statement**

Ultrafast imaging of chemical/physical reaction dynamics at nanoscale interfaces of a composite nanostructure requires resolutions in both space and time. Using single-pulse methodology, we directly and visually capture the irreversible eutectic related phase reactions of a single, same metal/semiconductor nanowire at nanometer-nanosecond spatiotemporal resolution by 4D electron microscopy. With a non-destructive free-standing sample preparation free from environmental disturbance that is important for statistical investigation, we have both qualitatively and quantitatively elucidated the transient phase reactions, and obtained important physical properties of the newly formed phases, such as latent heat and specific heat. Our work provides an efficient way of quantitatively determining physical properties of a nanoscale object with a tiny small quantity, especially when not available in bulk counterparts.

**Main text:**

A eutectic system refers to a unique thermodynamic entity that forms a lattice structure with a specific atomic ratio between the constituents. The eutectic temperature describes the lowest temperature at which the mixed substances become fully molten. Comprehensive understanding of eutectic related behavior offers a great potential for a wide variety of applications, including alloys for structural components and eutectic soldering in airplanes and cars (1), functional electronics in solar energy harvesting (2, 3) and eutectic bonding in chips for integrated circuits (4). Fundamental properties of solid microstructures from various eutectic reactions in complex systems depend on several factors, for example, the temperature that governs each phase reaction/transformation and interfacial nucleation of solid solutions (5, 6).

Eutectic related phenomena have usually been studied via post-processing measurements, and recently visualized by in situ techniques to investigate long time diffusion processes (7-13). Elucidation of the chemical/physical mechanism underlying those eutectic related processes requires time-resolved measurements of the dynamic phase formation. However, probing eutectic related phase reactions in nanostructures is very difficult because of the small sizes and the very short time scales involved, as well as the sensitivity issues to detect those transient processes that occur in far-from-equilibrium conditions for nanomaterials in a small quantity.

Recent development of 4D electron microscopy (4D EM) offers direct visualization of the transient behavior of individual nanostructures with high spatiotemporal resolution (14-19). Here, we report the direct observation and control of the irreversible phase reaction dynamics in free-standing Au/GaAs nanowires (NWs) by 4D EM with nanometer-nanosecond spatiotemporal resolution (see Fig. 1). GaAs NWs were chosen because III-V materials possess unique properties, such as direct bandgap, high carrier mobility, the possibility of bandstructure engineering using compositional tuning and hybrid of heterostructures, which demonstrate great potential for applications in optoelectronic and electronic devices (e.g., solar cells, sensors and transistors) (3, 20). It is fundamentally important to investigate how these materials behave in harsh environments (e.g., laser pulse bombardment for simulating the radiation effect) for understanding their stability. Furthermore, since eutectic related phenomena have been extensively studied in binary systems (e.g., Au-Si) under equilibrium conditions (7, 10, 11), the choice of a ternary system (Au-Ga-As) with pulsed laser excitation in 4D EM offers a window of opportunity for understanding eutectic dynamics in complex systems under non-equilibrium environment.

The eutectic related phase reaction processes of the NWs are induced and controlled by a temperature jump triggered by a laser heating pulse, followed by a delayed photogenerated electron packet to probe the spatiotemporal evolution of the nanostructures. Using single-shot imaging one can observe the transient intermediates of the NW during the reactions. Although the raised temperatures are below the melting points for the Au bead and the underneath GaAs NW, the observed reduction in NW length and the growth of the bead size are a result of the phase reactions between the two. For the time scale, the transient nucleation and growth of the formed phases are complete within ~80 ns while the morphological changes of the top bead lasts until 100 ns. The observed huge reduction rate of the NW length, which is not a simple diffusion-dominated process, will be discussed. Based on quantitative analysis and theoretical modeling of these measurements, we are able to extract useful thermodynamic parameters of the newly formed alloy phases determined by Bragg diffraction studies, leading to a better understanding of nanoscale phase reactions.

**Results and Discussion**

**Experimental implementation.** The difficulty of unambiguously exploring the ultrafast eutectic dynamics of a single Au/GaAs NW lies in isolating the NW from any transmission electron microscopy (TEM) supporting films (e.g., carbon films) while achieving a controllable NW geometry that is accurate enough for data analysis. We overcame this difficulty with a scheme of detecting the NWs that were grown perpendicular to the substrate surface. Namely, by directly putting the as-grown NWs still intact on the GaAs substrate (see Materials and Methods, and Fig. S1) into the 4D-EM (Figs. 1A and 1B), one could exclude any sample contamination or possible ambiguity caused by the holding films usually present for normal TEM grids. This type of sample preparation is simple and non-destructive with the freedom of investigating orientation-dependent (anisotropic) properties of the samples in their native environment. Because NWs grown on a substrate have similar orientations, better statistics are achievable through choosing a number of NWs for investigation. In addition, it becomes simple for quantitative elucidation of the phenomena without a need to take into account the effect from any supporting films (e.g., heat transfer and dissipation from the films).

The GaAs substrate was carefully aligned to be parallel to the electron beam so that the free-standing NWs were visible (otherwise, a very slight misalignment or tilting of the substrate, e.g., only 0.5º, would block the electron beam for the probing of NWs since the NWs were only hundreds of nanometers long in the direction vertical to the substrate surface). To facilitate imaging reaction dynamics from a single NW, the NWs were purposely grown on the substrate with large separation (several microns away from each other), as shown in the projections of the NWs (Fig. 1C). With this large separation among the NWs, the disturbance from any neighboring NW could be effectively avoided during the experiments. The transient morphological changes of the NWs during the eutectic reactions were captured by single-pulse imaging mode of 4D EM (see Materials and Methods).

**Morphology and structure of the NW.** Fig. 2 shows the typical micrographs and diffraction patterns of the Au/GaAs NWs in stroboscopic mode without the incident pump laser. In the images, the NWs are aligned vertically with respect to the GaAs substrate. A gold nanoparticle is located at the tip of each GaAs

NW. The targeted NW has a diameter of 108 nm, and its diffraction pattern (SA-1 region) is presented in the upper rightmost column. The pattern indicates that GaAs has a hexagonal wurtzite (WZ) structure with the growth direction oriented along the [000$\bar{1}$] direction. Shown in the inset of the leftmost middle column is a typical microstructure of the NW tip. The diffraction pattern of the tip, corresponding to the SA-2 region, exhibits additional diffraction spots. The unmarked spots are the same as that in the upper diffraction pattern, which come from the WZ GaAs. The marked extra <111> ones are from the cubic zinc-blende (ZB) segments of GaAs (21). This double-spot pattern indicates a twinned ZB component. The small ZB constriction (the dark line marked by an arrow is the ZB/WZ interface) is due to the growth termination procedure (20).

We performed further experiments in tilting conditions to verify how the NWs behaved during laser excitation (the bottom row of micrographs). It demonstrates that the NWs are straight (without bending) when viewed at different tilting angles. No breakage of the NW was observed either. This experiment ensured unambiguous determination of the shrinkage of the NW length after each single-pulse laser excitation.

**Imaging of eutectic related phase reactions.** To reveal the irreversible phase reactions, single-pulse imaging methodology was employed. The results of the size changes of the NW are shown in Fig. 3. To obtain a clear comparison of the NW at different excitation stages, for each experiment we enhanced the contrast between the top bead and the GaAs NW by taking two reference images with exposure to 100 electron pulses (Fig. S2). By comparing the single-pulse images with the corresponding reference images, the length of the NW at each stage can be determined with the guide of a ruler. Initially, the GaAs NW (without the top bead) has a length of 822 nm (0 laser shot). After a laser pulse at a fluence of 5.5 mJ/cm$^2$, the length of the NW became 797 nm, corresponding to a ~3% shrinkage. Compared to the initial GaAs

NW, there was a 40% length reduction after 23 laser pulses. The GaAs NW length (after 46 laser pulses) was finally reduced to 235 nm after the entire run of the experiment, which was associated with a total length shrinkage of about 70%. During the whole sequence of laser heating experiments, the choice of the number of pulses and laser fluence (from 5.5 to 40 mJ/cm$^2$) was set so that the NW could continue its shrinkage.

Although the NW axial length decreased significantly (Fig. 3, top), no obvious width change of the NW was observed (Fig. S3). Only the top bead, namely, the region with darker contrast than the pure GaAs part in the images, increased in size. Further analysis on the NW volume change revealed that the reduction of GaAs volume resulted in the increase of bead volume (see Fig. S4 for the calculations). These results indicate that the GaAs body remained solid phase while the vanished portion of the GaAs reacted with the top bead to form new phases during laser excitation.

**Identification of the phase formation.** Selected-area diffraction studies disclosed the nature of the newly formed phases during the reactions. We performed the electron diffraction in three modes: TEM (high electron counts), stroboscopic (1 kHz) and single-pulse modes (Fig. S5 and the associated discussion). For clearly demonstrating the diffraction spots, exposure times of 5 s in TEM mode and 120 s in stroboscopic mode were chosen. For diffraction in single-pulse mode (electron pulse duration of 10 ns), the extremely low electron dose (about five orders of magnitude lower than that in the stroboscopic mode with an exposure time of 120 s) made it inaccessible for disclosing the irreversible transient diffraction patterns of the newly formed phases with small quantity at the nanometer-nanosecond spatiotemporal resolution. Therefore, the structures of the alloy phases were identified in the TEM mode (Fig. S6 and Table S1).

Compared to the initial components (Fig. 2), additional diffraction spots from $Au_7Ga_2$ and AuGa phases appeared after the first few laser shots. According to the Au-Ga phase diagram (Fig. S7), the temperature for $Au_7Ga_2$ phase (22) is much lower than the melting point of either Au (1337 K) or GaAs (1511 K). As more GaAs was incorporated into the top bead, eutectic phases of AuGa and $AuGa_2$ were formed. The temperature needed for this eutectic reaction, liquid → AuGa + $AuGa_2$, is 725 K, still much lower than the melting points of the individual components in the initial materials (1337 K for Au and 1511 K for GaAs). It is noteworthy that we only detected Au-Ga related phases despite the possible existence of Au-Ga-As complex according to their ternary phase diagram (23, 24). Our observations indicate that in comparison with the Au-As complex, it is more favorable for the Au-Ga alloy formation during the ultrafast non-equilibrium reaction process. The solubility of As in Au is low, and the As species may vaporize or be removed from the interface during laser heating (25-28). The removal of the As species would lead to the slight reduction of the total volume, which was confirmed by the experimental observation in Fig. 3. We also note from Fig. S6 that new cubic ZB segments appeared after laser pulse excitation. The length of the ZB segments was estimated to be tens of nanometers by electron diffractions at different locations of the NW with a small aperture. This result indicates that with an incoming laser pulse the initial ZB component (Fig. 2) was alloyed with the top nanoparticle, and after laser excitation a small amount of new ZB segments appeared again. It suggests that compared to hexagonal WZ components, cubic ZB (normal bulk structure of GaAs) segments are preferable to precipitate near the interfacial region during non-equilibrium processes.

Calculations using two-temperature model for Au and three-temperature model for GaAs show that the lattice temperatures (at a fluence of 5.5 mJ/cm$^2$) of the top bead and GaAs body are 566 and 574 K (see Fig. S8 and Table S2), respectively, which triggers the reaction to form $Au_7Ga_2$ phase. At higher fluences, for example, 19.5 mJ/cm$^2$, the effective temperature of the top bead is 1196 K, high enough to facilitate the eutectic reactions for the formation of AuGa and AuGa$_2$ phases. These simulated results are in reasonably good agreement with our experimental observations.

**Thermal properties of the newly formed phases.** Theoretical modeling was employed to calculate the absorbed heat for both the bead and the NW so that the thermal properties for alloy phases during eutectic reactions could be determined (Supporting Information). Useful parameters, such as the latent heat and specific heat for the newly formed phases, are retrieved and the results are summarized in Table 1. We found that the latent heat of Au$_7$Ga$_2$ and AuGa are 8 and 21 kJ/mol while their specific heat are 62 and 41 J/mol.K, respectively. In comparison, for the initial phases Au and GaAs, the latent heat are 12 and 106 kJ/mol, and the specific heat are 25 and 48 J/mol.K, respectively (29, 30). Because of the presence of latent heat and the lower resultant bead temperature due to increasingly bigger bead size, one needs to increase laser fluences in order to shrink the NW further. Moreover, it is noteworthy that at high fluences (23 mJ/cm$^2$ or above) the GaAs NW has not melted even though the calculated temperatures (without taking into account the latent heat for melting) are higher than its melting point (Fig. 3). This is because the energy from a laser pulse is not sufficient to overcome the large latent heat of GaAs for its melting.

**Time-resolved dynamics.** To reveal the ultrafast eutectic dynamics, two sets of time-resolved experiments were performed on the NWs. Using stroboscopic approach (Fig. 4, top), the integrated diffraction intensity at a low fluence of 3 mJ/cm$^2$ is plotted as a function of time for the time constant determination of the non-equilibrium thermodynamics of the NW. As a result of the temperature jump initiated by the heating laser pulses, the integrated diffraction intensity shows a quick decrease after time zero and then a relatively slow recovery of the diffraction intensity. The cooling time constant $\tau$ was estimated to be 123±12 ns from the fitting. Aided by the extracted time scale, single-pulse imaging at that range was employed to capture the ultrafast eutectic process of the NWs (Fig. 4, bottom). Three rows of the single-shot images display the NW transient states after a single optical pulse (laser fluence of 19.5 mJ/cm$^2$) at different delay times. The first, second and third column of the images was taken at the stages of before, at specific delays (20-100 ns), and after the process ended, respectively. It is noticed from the first-row images that besides a length shrinkage between the left (before) and middle (at 20 ns delay) image, a clear shape change of the top bead was observed (marked with a circle). The shape of the bead continued its change even after the incident laser pulse has been removed (right image). When the delay time was 80 ns (second row), similar behavior, namely, the length shrinkage as well as the shape change, was also seen. However, the length and shape of the NW almost remained unchanged (third row) when observed at a delay time of 100 ns (middle) and at a state after the process ended (right). This observation indicates that the thermal energy induced by the laser heating pulse were insufficient to induce further obvious morphological change of the NW after a time of ~100 ns.

As known, heating of composite materials by ultrafast laser pulses usually involves very large thermal gradients near the interface. With the incoming heating pulses, the top bead and the GaAs NW are "hot" but remain solid until regions of melt phase nucleate. Under a large thermal gradient it enables fast motion of the involved atoms of the melt phase near the interface. The melt phase starts to nucleate at the Au/GaAs interface and to expand into the materials. Nevertheless, thermal diffusion near the interface limits the expansion of the melt region through cooling the photoexcited region. Therefore, after a short period of time, the temperature of the melt portion drops to the eutectic isotherm and the alloy phases begin to nucleate and finally solidify to form the alloys.

For the scenario of a free-space diffusion process, we estimate the diffusion constant using $l = \sqrt{2D\tau}$, where $l$ is the diffusion length, $D$ is the diffusion coefficient and $\tau$ is the time. From the single-pulse images in Fig. 4, the NW length showed a ~30 nm reduction at 20 ns (middle image in the first row) and continued a shrinkage of about 15 nm after the laser pulse (right image). The corresponding diffusion constant was estimated on an order of $10^{-4}$ $cm^2/s$, which is at least several orders of magnitude higher than the diffusivity of Au in GaAs and/or that of Ga in Au (28, 31-33). In comparison, the distance travelled by a ballistic gold or gallium atom with thermal velocity at the same temperature and time span would be on the order of microns, much larger than our observed NW length reduction. Therefore, the large reduction rate of the NW reported in this work suggests that the process cannot simply be diffusion-dominated processes, but rather indicates an important role due to the surface tension forces. Due to such surface-tension resultant forces, the reduction of 30 nm in NW length after a laser heating pulse could be fast, several orders of magnitude faster than the free-space diffusion or gravitational pull. The shape of the bead has maintained close to a sphere even though its size grows with number of laser heating pulses. Such spherical shape preservation is also an effect of surface tension, otherwise, the melt bead would drip down the GaAs NW similar to melt wax tearing down from a burning candle stick. Furthermore, there was no obvious length change of the NW from 80 ns (middle image in the second row) to the end state (right image). Such observation provides an important clue about the time needed for eutectic related phase reactions and solidification to complete. It indicates that the nucleation and growth of the melt phases were complete in ~80 ns, and after that they began to solidify due to the cooling.

**Polarization-dependent phenomena.** The above experiments were conducted under the condition where the electric field (*E*-field) of the pump laser pulse was parallel to the long axis of the NW. We have also examined the effect of *E*-field polarization on NW dynamics. Figure 5 shows polarization-dependent eutectic dynamics of individual NWs with the *E*-field of the laser pulse polarized either parallel or perpendicular to the NW axial direction (schematically illustrated in the inset of Fig. 5C). For parallel polarization, the morphology of the top bead (marked by a circle) began to change at a laser fluence of 6.5

mJ/cm$^2$ (Fig. 5A, left and middle columns). The threshold laser fluence for the morphological change of the bead, however, increased to 8.2 mJ/cm$^2$ with a perpendicular polarization (Fig. 5B, left and middle columns). It indicates the enhancement of eutectic reactions at the Au/GaAs interface for the parallel *E*-field polarization. With an incoming single laser pulse above the threshold (fluence of 13 mJ/cm$^2$), the length reduction of the GaAs NW was ~76 nm for parallel *E*-field polarization, compared to that of a smaller reduction of ~48 nm for perpendicular polarization (Figs. 5A and 5B, right column). Shown in Fig. 5C is the length reduction of the NW at a fluence of 13 mJ/cm$^2$ with a chain of laser shots (up to five). The NW length reduction becomes smaller with increasing the number of laser shots, and finally decreases to nearly zero on the fifth laser shot. The shrinkage of the NW length becomes increasing difficult (only increasing the number of laser shots) at a fixed laser fluence due to the presence of latent heat of the newly formed alloys.

It is noteworthy that at each laser shot the length reduction of the NW is always larger in parallel *E*-field polarization than that in the perpendicular one (Fig. 5C). This result elucidates the observed reaction is dependent on the laser *E*-field polarization. Under laser pulse excitation, collective electron oscillations on a nanostructure by incident laser (electromagnetic field) gives rise to localized surface plasmons (strong charge densities of opposite sign at both sides of the nanostructure), whose field distribution is determined by the *E*-field polarization of the incident laser. Compared to perpendicular *E*-field polarization, the localized surface plasmon fields in the case of parallel polarization (electron oscillations are along the long axis of the nanostructure) will be enhanced near the bead/NW interfacial region (34-38), resulting in strong local photothermal effect at the interface where the eutectic reaction takes place. In addition, parallel *E*-field polarization leads to a further enhancement in optical absorption (causing higher temperature) along the NW (39, 40). Because of these enhancement effects, the laser fluence threshold for achieving the eutectic reactions reduces in the case of parallel field polarization; while at the same laser fluence the length reduction of the NW (swallowed by the top nanoparticle due to eutectic reactions) is larger in the parallel polarization than that in the perpendicular one, as experimentally observed in Fig. 5.

**Conclusions**

In summary, our findings suggest that the eutectic related phase reaction processes of the NWs can be controlled/designed via tailoring the fluence, polarization and number of laser pulses in the pulse sequence, providing the potential of eutectic bonding/welding to a desired/targeted location in micro/nanodevices. The ability to fine tune the nanoparticle size and shape is important for plasmonics applications (41-43). Additionally, using an easy design and non-destructive preparation technique of as-grown free-standing nanowires (without supporting films), we are able to effectively study the anisotropic (orientation-dependent) properties of the NWs in their native environment with better statistics. Using 4D single-pulse imaging of multi-phase eutectic related processes in nanostructured materials as an example, we clearly demonstrate the potential of this novel technique for exploring irreversible chemical/physical reaction dynamics at high spatial and temporal resolution so that the underlying mechanisms in complex systems could be unraveled. The powerful capabilities of 4D EM provide in-situ manipulation and visualization of irreversible processes in space and time, and therefore offers a valuable tool to investigate photochemical reactions as well as macromolecular dynamics in biological systems.

**Materials and Methods**

**Preparation of Au/GaAs NWs.** The gold-catalyzed GaAs NWs were epitaxially grown on a GaAs $(111)_B$-oriented substrate by metal-organic chemical vapor deposition. The catalysts used for the growth were gold nanoparticles. Initially, the GaAs substrate was treated with poly-L-lysine solution followed by a solution of colloidal Au nanoparticles ~100 nm in diameter. The major role of catalyst Au is to lower the nucleation energy of each grown layer at the seed/substrate interface, promoting the growth of NWs. Prior to growth, the substrate was annealed at 600 °C under $AsH_3$ ambient to remove surface contaminants. The substrate was then cooled to 550 °C, and the NW growth proceeded by injecting the precursors of trimethylgallium and $AsH_3$ to the growth chamber with ultra-high purity hydrogen as the carrier gas. Details of the growth procedures of GaAs NWs are available elsewhere (44, 45).

**Sample preparation for 4D imaging.** Normally, to prepare NW samples for transmission electron microscopy studies, the NWs grown on a substrate are transferred to a metal grid (200-2000 mesh) or to a grid with supporting films (carbon, silicon monoxide, silicon nitride etc.). During the transfer process, the NWs are either broken into several segments (permanent damage), contaminated or aggregated. In this work, we prepared the NW samples without any damage or using supporting films. First, the substrate containing free-standing GaAs NWs was cut into small pieces with dimensions of ~2×1×0.3 $mm^3$. One side of the small piece (the side of the 0.3 mm-thickness) was glued on a common copper O-ring. Then the copper O-ring with the sample was directly mounted on the EM holder for investigation (see Fig. S1). This method of sample preparation excludes any disturbance from the supporting films normally present in EM grids. Furthermore, this type of sample preparation allows us to investigate the properties of a single NW with better statistics (because all NWs grown on the substrate have the similar orientation and direction, we can choose many of them to obtain the statistical data).

**Spatiotemporal visualization of eutectic dynamics.** The morphology of GaAs NWs was investigated using an FEI scanning electron microscope. The studies of the eutectic dynamics and control of Au/GaAs NWs were conducted by 4D-EM (Fig. 1A). The 4D-EM is equipped with laser systems, which can operate in either stroboscopic (for reversible processes) or single-pulse mode (for irreversible dynamics). Both modes were used in this work. Picosecond green laser pulses at 532 nm (16 ps pulse duration) were used to heat the sample and to initiate the reactions, whereas nanosecond ultraviolet laser pulses at 266 nm (10 ns pulse duration) were directed to the photocathode to generate electron pulses. The photoelectrons in each pulse for probing were then accelerated to 120 keV, corresponding to a de Broglie wavelength of 3.3 pm. The timing between the pump pulse and the probe pulse was controlled by changing the delay time between the two through a digital delay generator.

The NW transient morphologies during the eutectic reactions were captured by single-pulse imaging mode of 4D-EM, where an entire image was acquired with only one electron pulse (contains ~$10^5$ electrons) following an excitation laser pulse. For stroboscopic experiments the repetition rate was set at 1 kHz, ensuring the full recovery of the dynamical process. Details of the procedure are given in previous studies from this group (46-48). Selected-area electron diffraction was performed to identify the structures of the NWs and the newly formed alloy phases. Three kinds of modes were used, namely, TEM, stroboscopic (1 kHz) and single-pulse modes.

**Polarization-dependent behavior.** We used a femtosecond laser (wavelength of 520 nm, pulse duration of 350 fs) to excite the NW dynamics, and probed the eutectic behavior of the NWs with an electron beam. A polarizer was used to set the *E*-field polarization of the laser pulse either parallel or perpendicular to the long axis of the NWs (see schematic diagram in Fig. 5). For each polarization direction, a single laser pulse irradiated the NW sample at each time. The subsequent morphological change and length reduction of the NWs, which were due to the reaction between the Au and GaAs, were directly visualized by the electron beam.

**Acknowledgements**

We acknowledge the financial support from the Air Force Office of Scientific Research (Grant No. FA9550-11-1-0055S) in the Gordon and Betty Moore Foundation for Physical Biology Center for Ultrafast Science and Technology at California Institute of Technology. The Australian authors thank the Australian Research Council for the financial support and the Australian National Fabrication Facility for access to the epitaxial facilities used in this work. We appreciate J. S. Baskin and J. S. Huang for very helpful discussion.

**Author contributions**

B.C. and A.H.Z. conceived the experiment, B.C. and X.W.F. designed and conducted the experiment, J.T. performed theoretical modeling and data analysis, M.L., H.H.T. and C.J. grew the samples, all authors analyzed the data and wrote the paper.

**Table 1. Latent heat and specific heat for two alloy phases.**

| Phase | Latent heat (kJ/mol) | Specific heat (J/mol.K) |
|---|---|---|
| $Au_7Ga_2$ | 8 | 62 |
| AuGa | 21 | 41 |

**Figure captions**

**Fig. 1. 4D electron microscopy for imaging phase reactions in a free-standing NW.** (**A**) Picosecond green optical pulse enables the eutectic reaction of Au/GaAs NWs to be probed by a delayed electron pulse (generated by a nanosecond ultraviolet laser pulse) at a given time delay. (**B**) Scheme of free-standing NWs on substrate integrated in 4D-EM (see Fig. S1). The laser heating pulse is incident upon the NW at a nearly vertical angle while a series of eutectic events are captured by single-pulse methodology. (**c**) Projections of the NWs. The NWs were intentionally grown far apart from each other to exclude mutual perturbation during the measurement.

**Fig. 2. Images, diffractions and tomography micrographs of the Au/GaAs NWs.** Top row: Stroboscopic image of the NW and its diffraction pattern (marked circle in the image). The diffraction pattern indicates a hexagonal WZ structure of GaAs. Middle row: Image and diffraction pattern of the tip region. Additional diffraction spots appeared besides the same ones as that in the top row. These extra spots are from the cubic ZB segments, which can be seen from an enlarged image in the inset of the left image (scale bar, 50 nm). Bottom row: Images of the NW at different tilting angles ranging from -10º to 50º. The dotted lines indicate the length measured from the tip to the reference point (marked by an arrow). At all angles the NW is straight, suggesting that neither inclination nor breakage of the NW occurred after the laser pulse excitation.

**Fig. 3. Single-pulse imaging of transient changes of the NW, and the calculated temperatures.** Top: Single-pulse imaging of the NW length reduction due to eutectic reactions. The NW shrank along its axial direction while its lateral width had no obvious change. About 70% length reduction of the NW (without the top bead) was observed after a total of 46 laser shots. The rulers (white lines) indicate the NW length. The electric field polarization of the laser pulse was parallel to the long axis of the NW. Bottom left:

Volume change of the NW. The volume of GaAs NW reduces while that of top bead increases. The total volume slightly decreases. Bottom right: Calculated temperatures at different laser fluences. The temperature of Au or GaAs bead was estimated by assuming 100% of the component while the effective temperature of the bead was calculated by considering the Au:GaAs ratio in the bead. The dotted line represents the lowest temperature to trigger the reaction. Because of the large latent heat of GaAs, the GaAs NW has not melted though the temperatures (without taking account of the latent heat) are higher than its melting point (dashed line) at high fluences. The solid red line indicates the actual temperature of the GaAs NW when its latent heat is considered.

**Fig. 4. Time-resolved ultrafast dynamics of the NW.** Top: Normalized diffraction intensity vs. delay time, indicating cooling dynamics with a time constant of 123±12 ns. The fitting is shown in the red line. Bottom: single-shot images of the NW at a specific delay time. The images in the left, middle and right columns correspond to the states of the NW before, at specific delays, and after the process ended (the incident laser pulse has been removed), respectively. Three rows of images are associated with the NW states at the delay time of 20, 80 and 100 ns, respectively. Besides the length shrinkage, the transient shape changes of the top bead (marked with a circle) at different delays were also captured. The white rulers indicate the NW length. The electric field of the pump laser pulse was parallel to the long axis of the NW.

**Fig. 5. Polarization-dependent dynamics.** (**A**) EM micrographs of a NW before and after a laser pulse whose *E*-field polarization is parallel to the long axis of the NW (scale bar, 100 nm). (**B**) EM micrographs of a NW before and after a laser pulse but with a perpendicular *E*-field polarization (scale bar, 100 nm). The middle and right columns in (**A**) and (**B**) correspond to the images at different laser fluences. The circles indicate the morphological change of the top bead. (**C**) Length reduction of the GaAs NWs at different polarization directions. For each polarization, the laser fluence is 13 mJ/cm$^2$. The length reduction was determined from the NW length change after each laser shot. Inset, schematic diagram

showing a NW excited by a laser pulse with its *E*-field polarization parallel or perpendicular to the long axis of the NW.

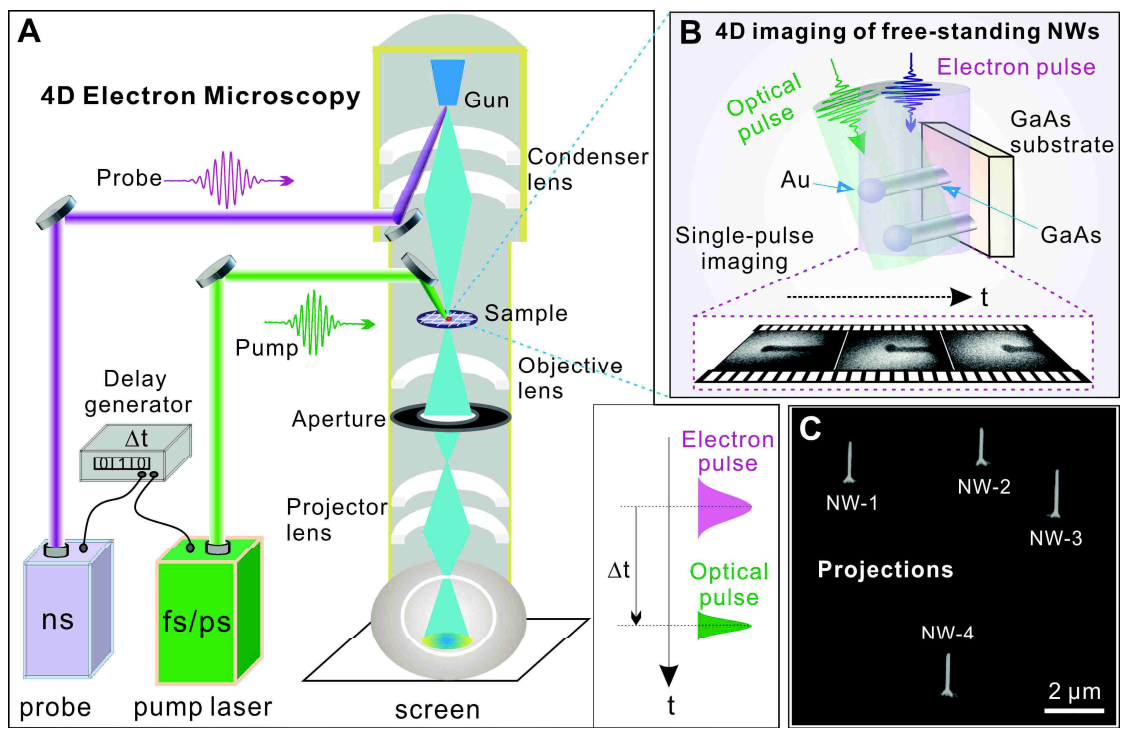

**Figure 1**

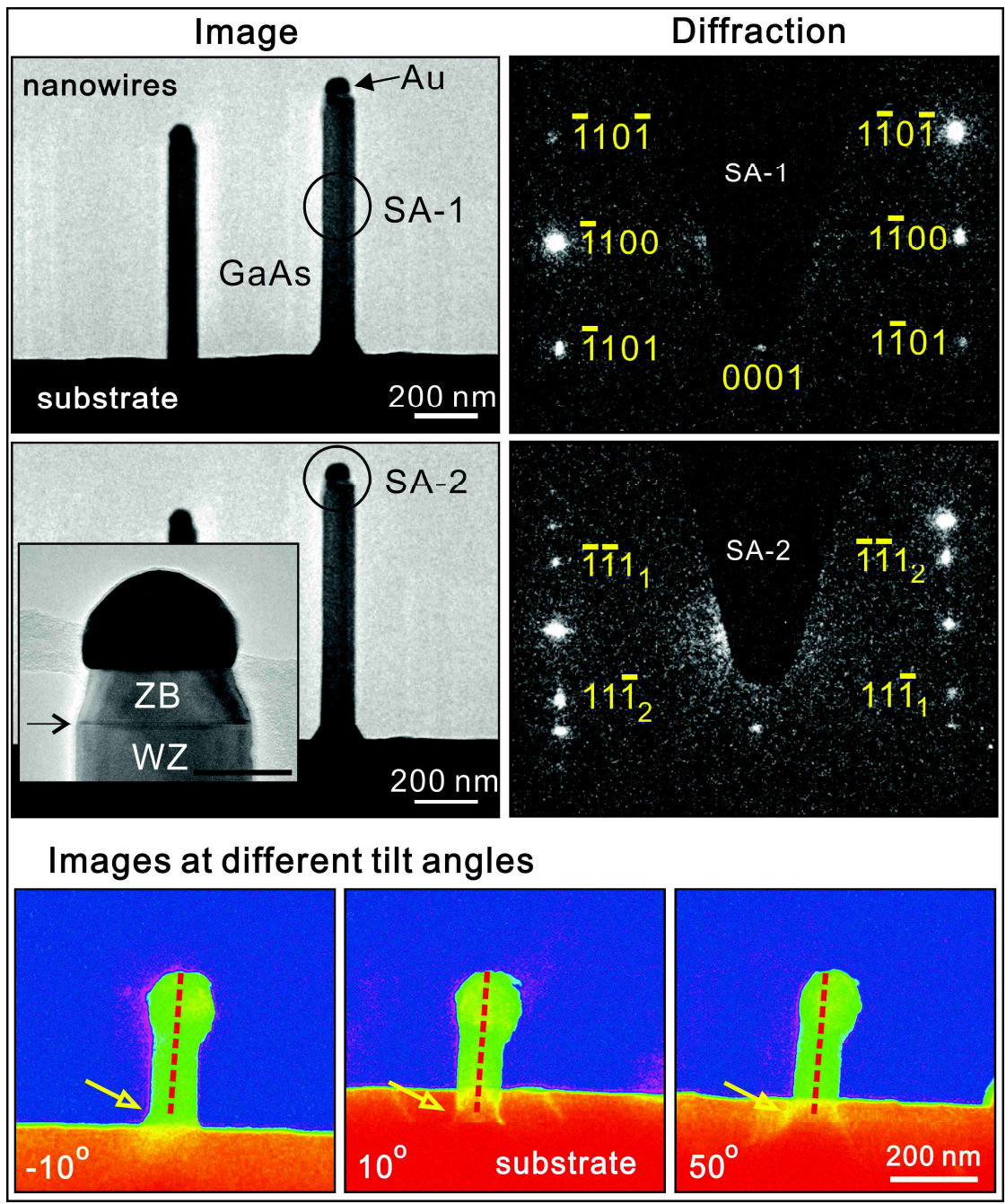

Figure 2

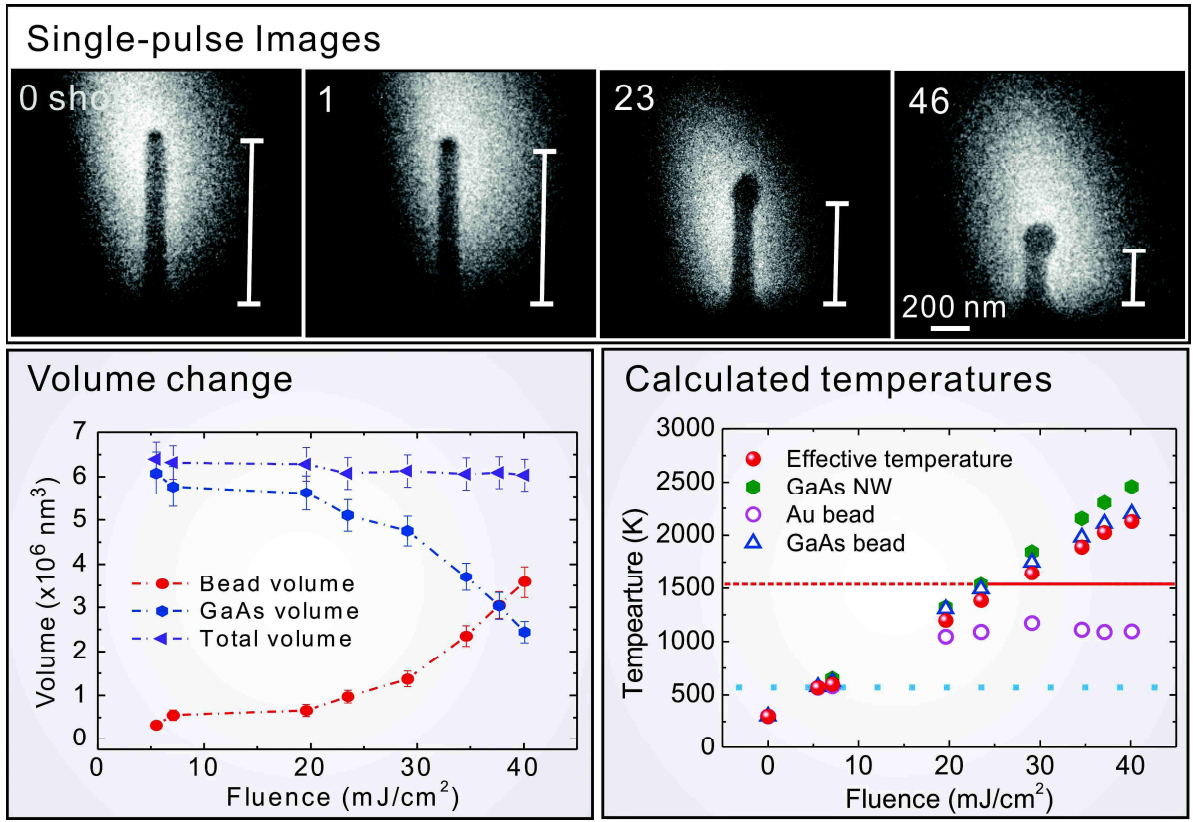

**Figure 3**

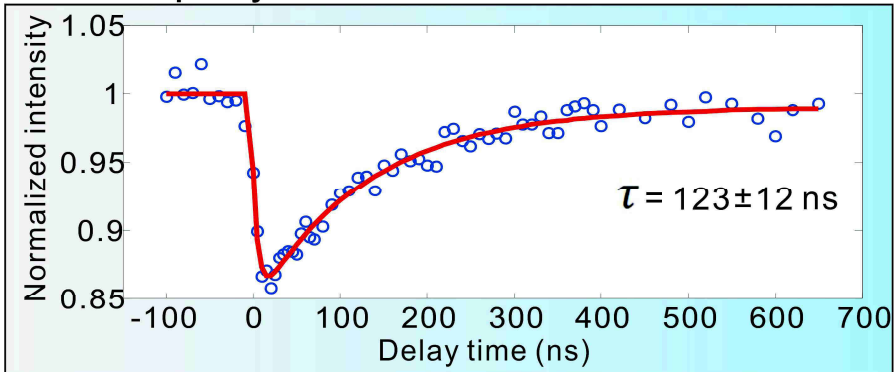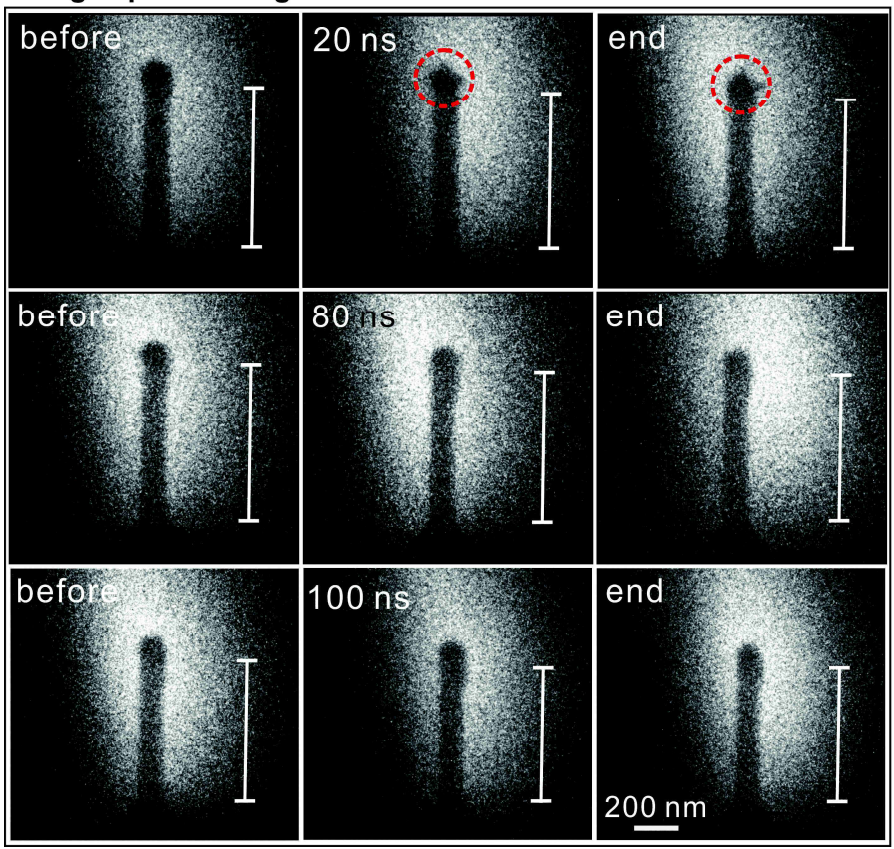

**Figure 4**

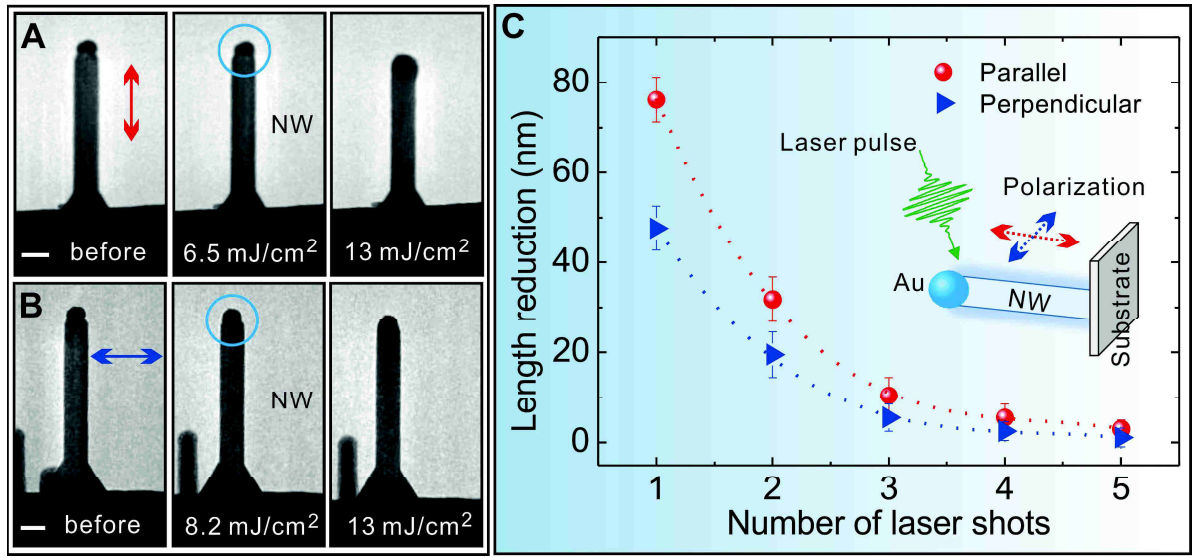

**Figure 5**

Supporting Information for

# Dynamics and control of gold-encapped gallium arsenide nanowires imaged by 4D electron microscopy

Bin Chen,[1,*] Xuewen Fu,[1] Jau Tang,[1,*] Mykhaylo Lysevych,[2] Hark Hoe Tan,[3] Chennupati Jagadish,[3] Ahmed H. Zewail[1,†]

[1]Physical Biology Center for Ultrafast Science and Technology, Arthur Amos Noyes Laboratory of Chemical Physics, California Institute of Technology, Pasadena, CA 91125, USA

[2]Australian National Fabrication Facility, Research School of Physics and Engineering, The Australian National University, Canberra, ACT 2601, Australia

[3]Department of Electronic Materials Engineering, Research School of Physics and Engineering, The Australian National University, Canberra, ACT 2601, Australia

*Corresponding author. Email: bchen5@caltech.edu, jautang@caltech.edu

[†]Deceased

The PDF file includes:

Supporting Notes

Figures S1 to S8

Tables S1 to S2

Supporting References

**Supporting Notes**

Sample preparation for electron microscopy imaging

The free-standing nanowires (NWs) still intact on the substrate were mounted on a double-tilt holder (Fig. S1A). A magnified image of the sample (marked by a rectangular box) is shown in Fig. S1B. The sample preparation is schematically illustrated in Fig. S1C. One thickness (~0.3 mm) side of the substrate (typical dimensions of ~2×1×0.3 mm$^3$) was glued on a Cu O-ring. Fig. S1D shows a typical transmission electron microscopy (TEM) image of the NWs (with substrate). The substrate was seen as completely dark contrast because no electron beam was transmitted.

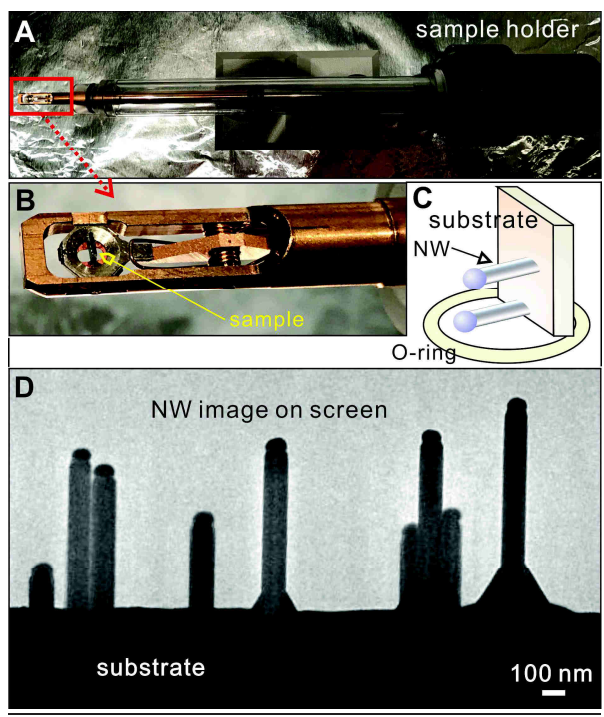

**Fig. S1.** Free-standing sample preparation for electron microscopy investigation. (**A**) Overview of a sample holder. (**B**) Specimen was mounted on the sample holder. This magnified image was from the box shown in (**A**). (**C**) Schematic diagram of the free-standing NW sample preparation. (**D**) TEM image of the NWs viewed by a charge-coupled device (CCD).

Length change of the NW

Transient morphological changes of the Au/GaAs NWs were imaged in the single-pulse mode. The lengths of the GaAs NW (without the top bead) before and after laser pulse excitation are marked by white rulers (Fig. S2). In order to enhance the contrast between the top bead and the GaAs NW, and to make a comparison of the NW at different excitation stages, two reference images were taken with exposure to 100 electron pulses. One was before the arrival of a laser pulse while the other one was after the whole process ended (immediately after an incoming laser pulse an intermediate state at each delay time was captured. Then the shutter of the pump laser was closed, and we took the final reference image. This final image means the one after the process ended). By such a way, the length of the GaAs NW at each stage was indicated by a ruler (Figs. 3 and 4 of the main text).

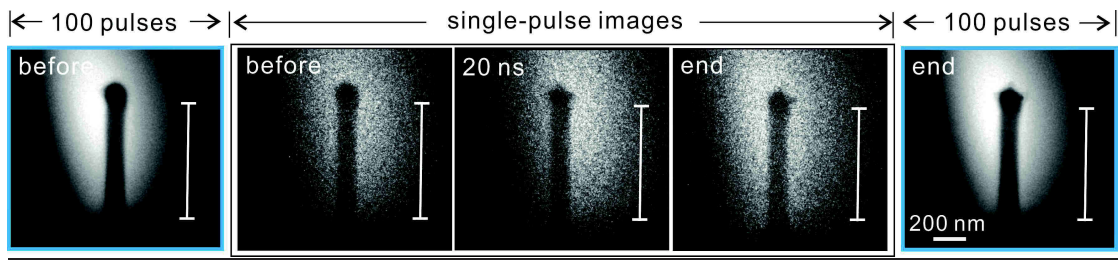

**Fig. S2.** Length reduction of the NW during laser pulse excitation. Intermediate state of the NW at a specific delay time was visualized by single-pulse method. To enhance the contrast and to make a clear comparison, two reference images were taken with 100 electron pulses at the stage before the excitation and after the process ended.

Width change of the NW

Fig. S3 shows a comparison between two TEM images of a NW. These two images correspond to the initial and final state of the NW, respectively. Because GaAs NW was incorporated into the top Au bead that enabled the eutectic reactions, the top bead became bigger than the initial one. However, it is noteworthy that the width of the NW body shows no obvious change.

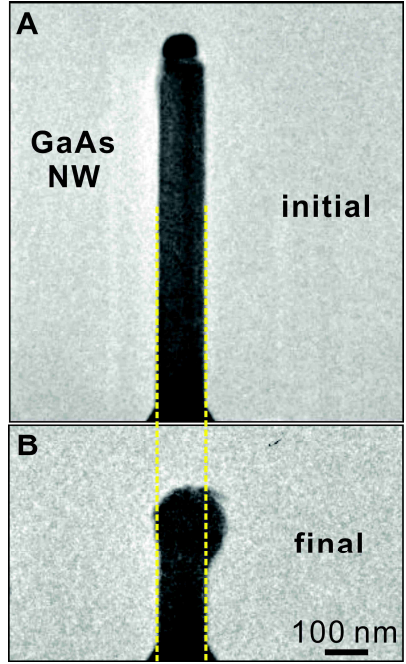

**Fig. S3.** TEM images of a NW, before (**A**) *vs*. after laser excitation (**B**). The dashed lines indicate the left and right edges of the NW.

Volume change of the NW

The volume change of the NW was estimated according to the schematics shown in Fig. S4. GaAs has a hexagonal shape in cross-sectional geometry (1) while the initial Au bead shows a segment shape of a sphere. The initial volume $V_0$ of the Au/GaAs is,

$$V_0 = S \times l + 4/3\, \pi R^3 - \pi(R-m)^2 \times (R - \frac{R-m}{3}) \tag{1}$$

where $l$ and $S$ are the length and cross-sectional area of GaAs, $R$ is the initial radius of the Au bead and $m$ is the height from the center point to the bottom of the sphere segment. Based on the hexagonal geometry, $S$ equals to $\frac{3\sqrt{3}}{2}d^2$, where $d$ is the half width of the GaAs NW.

The volume reduction of the GaAs NW as a result of the length shrinkage $\Delta l$ becomes,

$$\Delta V_1 = \Delta l \times S \tag{2}$$

The volume increase of the top bead with a final radius of $R'$ is given by,

$$\Delta V_2 = 4/3\, \pi R'^3 - [4/3\, \pi R^3 - \pi(R-m)^2 \times \left(R - \frac{R-m}{3}\right)] \tag{3}$$

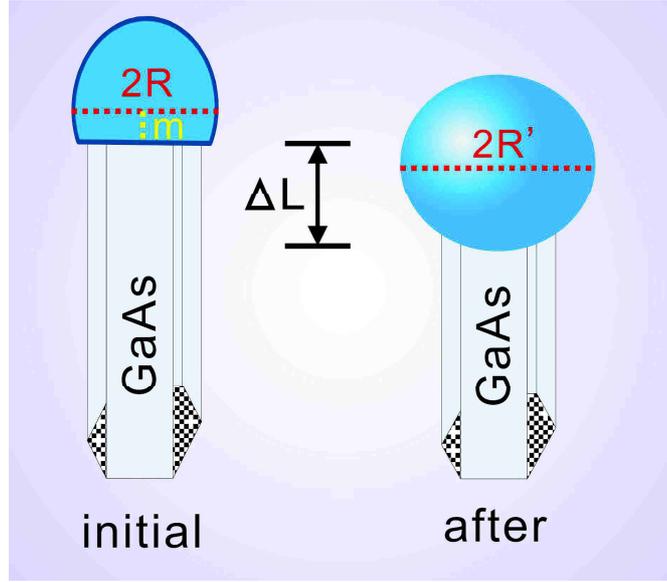

**Fig. S4.** Schematic diagram of the NW and the bead used for estimation of the volume changes. GaAs has a hexagonal cross-section with an area of S. The length change $\Delta l$ of GaAs was measured by the difference between the initial length and the one after a laser shot. The initial Au bead shows a segment of a sphere with a diameter and height of $2R$ and $(R+m)$, respectively. The top bead after laser excitation is assumed to be a sphere with a diameter of $2R'$.

Therefore, the total volume change of the Au/GaAs is written as,

$$\Delta V = \Delta V_1 + \Delta V_2 \tag{4}$$

Substituting the parameters $l$, $\Delta l$, $R$, $R'$, $m$, and $d$ measured from the TEM images into equations (1)-(4), the volume changes of the bead and the NW are plotted in Fig. 3 of the main text.

Newly formed alloy phases after laser excitation

We applied electron diffraction to examine the newly formed alloy phases. Fig. S5 show the electron microscopy (EM) micrograph and diffraction patterns of a NW took at different modes. A region marked by a circle in Fig. S5A was used for performing the electron diffraction patterns (Fig. S5B). Three modes, namely, TEM, stroboscopic (1 kHz) and single-pulse modes were used for comparison. When the exposure time was 5 s in the TEM mode, clear diffraction spots from the sample appeared. In the stroboscopic mode, similar diffraction spots were present when the exposure time was largely increased to 120 s (only several diffraction spots appeared at the exposure time of 10 s, but with much weaker intensity due to lower electron dose). In the single-pulse mode (only a single electron pulse), however, no clear diffraction spot was observed. The reason is shown below. There are only ~$10^5$ electrons in a single electron pulse of the single-pulse mode (effective exposure time of only 10 ns, which is the duration of the electron pulse). Compared to the stroboscopic mode (showing the clear diffraction spots), whose repletion rate and exposure time were 1 kHz and 120 s, respectively, the electron dose in the single-pulse mode was about five orders of magnitude lower. Note that increasing the exposure time in the single-pulse mode did not help because only a single electron pulse (10 ns duration) was present for probing the sample. So current 4D EM facility is not accessible to the irreversible transient diffraction studies (single-pulse mode) of the NWs with small quantity at the nanometer-nanosecond spatiotemporal resolution shown in this work. Therefore, the newly formed alloy phases were identified in the TEM mode shown in the following part.

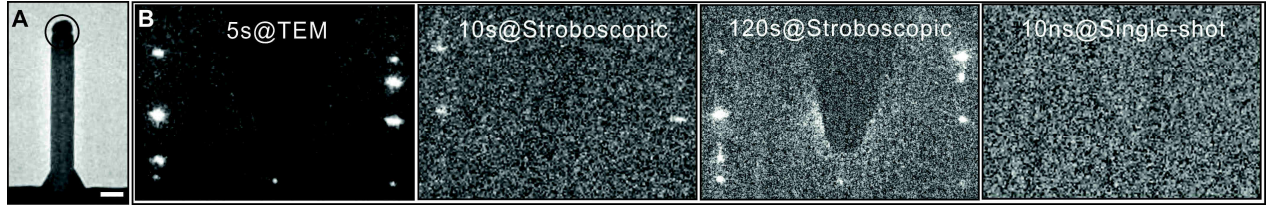

**Fig. S5.** Image and diffraction patterns from the top part of a NW. (**A**) TEM micrograph of the NW. Scale bar, 100 nm. (**B**) Diffraction patterns from a region of the NW marked by a circle. They were taken at three different modes: TEM, stroboscopic (1 kHz) and single-pulse modes. Different exposure times to the electron beam were applied, and the patterns at these exposure times were compared. For the single-pulse mode, the effective exposure time was only about 10 ns (duration of an electron pulse).

Fig. S6 show the diffraction patterns of the top part of a NW before and after the laser excitation. These images were taken under the off-axis conditions so that the diffraction intensities from the newly formed phases were enhanced. In the initial state (Fig. S6A), the diffraction spots were from GaAs, similar to that shown in Fig. 2 of the main text. After the first three laser pulses, however, additional diffraction spots were seen (Fig. S6B). These spots were from $Au_7Ga_2$ $\{11\bar{2}4\}$ ($Au_7Ga_2$-L, low-temperature phase) and AuGa $\{200\}$, respectively. The lattice constants used for the identification of these phases are listed in Table S1. In the final state (Fig. S6C), namely, as more GaAs component was incorporated into the top bead, new phases from $AuGa_2$ $\{111\}$ and Ga $\{131\}$ were detected besides the AuGa phase. Compared to the patterns in Fig. 2 of the main text, the diffraction spots from cubic zinc-blende (ZB) segments also appeared after laser excitation (indicated by the arrows). For the As element, the As species may vaporize or be removed from the interface during the laser heating (6, 7).

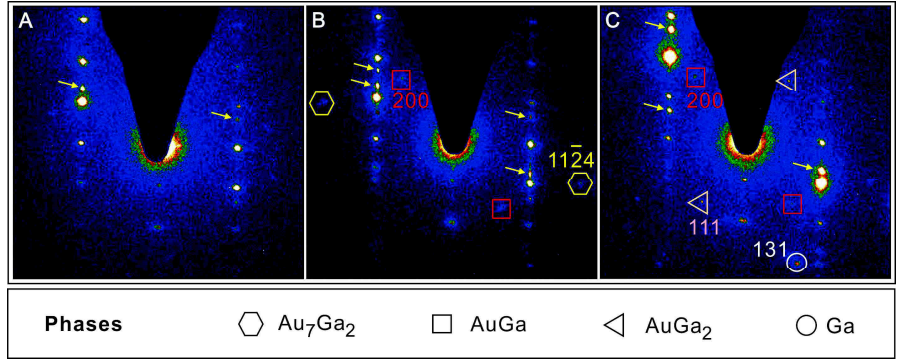

**Fig. S6.** Diffraction patterns from the top part of a NW. (**A**) Initial state without any laser pulse excitation. (**B**) After the first three laser pulses. Compared to the initial state in (**A**), additional diffraction spots from $Au_7Ga_2$ and AuGa were detected. (**C**) Final state after 46 laser pulses. AuGa, $AuGa_2$ and Ga phases were present. The hexagonal, rectangular, triangular and circular symbols represent the phases of $Au_7Ga_2$, AuGa, $AuGa_2$ and Ga, respectively. The arrows indicate the diffraction spots from the ZB segments of GaAs.

**Table S1. Lattice constants of the newly formed phases.**

| Phase | Structure | a (Å) | b (Å) | c (Å) | Reference |
|---|---|---|---|---|---|
| $Au_7Ga_2$ | Hexagonal | 7.721 | / | 8.751 | (2-5) |
| AuGa | Orthorhombic | 6.397 | 6.262 | 3.463 | |
| $AuGa_2$ | Cubic | 6.076 | / | / | |
| Ga | Orthorhombic | 4.519 | 7.657 | 4.526 | |

Phase diagram of Au-Ga binary system

From the diffraction analysis the newly formed phases in the top bead are identified as Au-Ga alloys. Fig. S7 presents the Au-Ga binary phase diagram (8). The melting points of Au and GaAs are 1337 and 1511 K, respectively. However, the temperature needed for the eutectic reaction can be much lower than the melting point of any pure component. For example, to obtain the phases of $Au_7Ga_2$ (low-temperature phase, $Au_7Ga_2$-LT) and AuGa, the temperatures are ~555 and 620 K, respectively. On the other hand, the temperature needed for the eutectic reaction of liquid → AuGa + $AuGa_2$ is about 725 K.

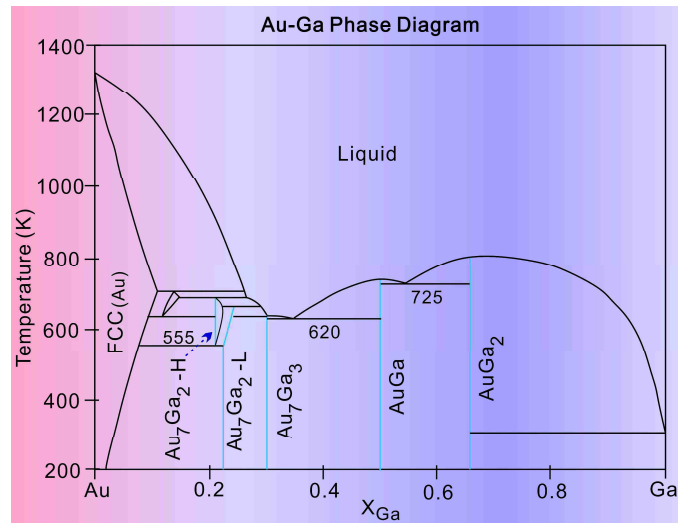

**Fig. S7.** Au-Ga binary phase diagram redrawn from ref. 8. The horizontal axis is the atomic fraction of Ga component.

Modeling of laser heating of gold and GaAs

Laser heating of a GaAs NW with a Au tip and the induced eutectic phase reaction can be modeled using the following theoretical approach. First, we consider laser heating of Au nanoparticle alone using the conventional two-temperature model (TTM) which deals with temporal and spatial temperature changes for both electron and phonon subsystems. Then, we treat laser heating of a GaAs NW using a three-temperature model (3TM) to deal with the temporal and spatial evolution of charged carrier density, carrier temperature, optical phonon temperature and acoustic phonon temperature. Because the optical reflectivity could be influenced by the nanoparticle size (Fig. S8), in our simulations for a given bead size we have incorporated empirical size-dependent reflectivity reported in literature (9-15). To calculate the lattice temperature for the bead with a mixture of Au and alloy, the following approach was used. We first calculated the lattice temperature for a bead of Au (assuming 100% Au) and also a bead of GaAs (100% GaAs), and then we estimated the effective lattice temperature based on their composition percentage within the bead. This approximation is a necessary step in the simulation because the actual optical and thermal parameters used in TTM and 3TM modeling for a composite nanoparticle have not been known in literature.

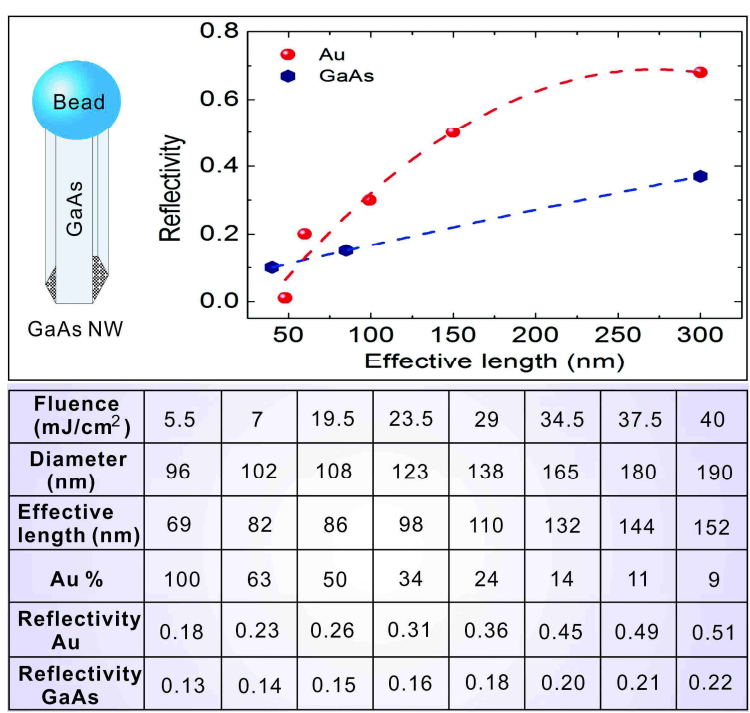

| Fluence (mJ/cm$^2$) | 5.5 | 7 | 19.5 | 23.5 | 29 | 34.5 | 37.5 | 40 |
|---|---|---|---|---|---|---|---|---|
| Diameter (nm) | 96 | 102 | 108 | 123 | 138 | 165 | 180 | 190 |
| Effective length (nm) | 69 | 82 | 86 | 98 | 110 | 132 | 144 | 152 |
| Au % | 100 | 63 | 50 | 34 | 24 | 14 | 11 | 9 |
| Reflectivity Au | 0.18 | 0.23 | 0.26 | 0.31 | 0.36 | 0.45 | 0.49 | 0.51 |
| Reflectivity GaAs | 0.13 | 0.14 | 0.15 | 0.16 | 0.18 | 0.20 | 0.21 | 0.22 |

**Fig. S8.** Top: The size-dependent reflectivity of Au and GaAs (9-15). Bottom: The reflectivity of Au and GaAs adjusted according to their actual size observed in the experiment. To simplify the theoretical modeling, a cube-shaped top bead for the corresponding spherical volume was assumed. The diameter of the bead was converted to the effective length of the cube. The gold percentage was estimated by the ratio of the initial gold volume to the final bead volume.

According to the TTM with a laser beam along the z-axis, the evolution of the electron temperature $T_e(z,t)$ and the lattice temperature $T_L(z,t)$ follows (16-18),

$$C_e \frac{\partial}{\partial t}T_e(z,t) = \frac{\partial}{\partial z}\left(k_e \frac{\partial}{\partial z}T_e(z,t)\right) - g[T_L(z,t) - T_e(z,t)] + S(z,t)$$

$$C_L \frac{\partial}{\partial t}T_L(z,t) = -g[T_e(z,t) - T_L(z,t)] \tag{5}$$

where $C_L$ is the specific heat for phonons, and $S(z,t)$ is the incident laser pulse profile. At a low laser excitation fluence, one usually assumes that the electron/phonon coupling $g$ is a constant, the specific heat $C_e$ for electrons has linear dependence on electron temperature, and the electronic thermal conductivity $k_e$ is approximated by $K_e T_e/T_L$. Such approximations have shown to be inaccurate for electron temperature beyond 3000 K. Here we use the following improved formulae for their temperature dependence,

$$k_e = \frac{4.26 \times 10^{13} T_e}{1.20 \times 10^7 T_e^2 + 1.23 \times 10^{11} T_L} \tag{6}$$

and Padé approximation for the electron/phonon $g$ and $C_e$, respectively,

$$g = 1.0 \times 10^{17} \frac{\sum_{n=0}^{4} A_1(n)(T_e/10^4)^n}{1+\sum_{n=1}^{4} A_2(n)(T_e/10^4)^n}$$

$$C_e = 1.0 \times 10^{17} \frac{\sum_{n=0}^{3} B_1(n)(T_e/10^4)^n}{1+\sum_{n=1}^{3} B_2(n)(T_e/10^4)^n} \tag{7}$$

where $A_1(0) = 0.257$, $A_1(1) = -0.549$, $A_1(2) = 0.553$, $A_1(3) = -0.650$, and $A_1(3) = 6.269$; $A_2(1) = -2.544$, $A_2(2) = 4.566$, $A_2(3) = -1.479$, and $A_2(4) = 2.541$; $B_1(0) = -0.043$, $B_1(1) = 8.451$, $B_1(2) = -28.797$, and

$B_1(3) = 68.387$; $B_2(1) = -1.645$, $B_2(2) = 2.539$, and $B_2(3) = 0.702$. All other relevant parameters for Au and GaAs are listed in Fig. S8.

For laser heating of a GaAs NW with an incident intensity profile $I(z,t)$, we employed three-temperature model (3TM) that describes time evolution of four subsystems, namely, the charge carrier density $N_C(z,t)$, and three subsystem temperatures such as carrier temperature, optical and acoustic phonon temperatures. Defining the internal energy for the charge carriers $U_C(z,t)$, and the internal energy for the longitudinal optical (LO) phonons $U_O(z,t)$ and the longitudinal acoustic (LA) phonons $U_A(z,t)$, one has (19-21),

$$\frac{\partial}{\partial t} N_C(z,t) = \frac{\alpha_1 I(z,t)}{\hbar \omega} - \gamma_A N_C^3(z,t)$$

$$\frac{\partial}{\partial t} U_C(z,t) = \frac{\partial}{\partial z}\left[k_C \frac{\partial}{\partial z} T_C(z,t)\right] + \alpha_1 I(z,t) - \frac{3k_B N_C(z,t)}{2}\left[\frac{T_C(z,t) - T_O(z,t)}{\tau_{C-O}}\right]$$

$$- \frac{3k_B N_C(z,t)}{2}\left[\frac{T_C(z,t) - T_A(z,t)}{\tau_{C-A}}\right]$$

$$\frac{\partial}{\partial t} U_O(z,t) = \frac{3k_B N_C(z,t)}{2}\left[\frac{T_C(z,t) - T_O(z,t)}{\tau_{C-O}}\right] - C_O\left[\frac{T_O(z,t) - T_A(z,t)}{\tau_{O-A}}\right]$$

$$\frac{\partial}{\partial t} U_A(z,t) = \frac{\partial}{\partial z}\left[k_A \frac{\partial}{\partial z} T_A(z,t)\right] + \frac{3k_B N_C(z,t)}{2}\left[\frac{T_C(z,t) - T_A(z,t)}{\tau_{C-A}}\right] + C_O\left[\frac{T_O(z,t) - T_A(z,t)}{\tau_{O-A}}\right] \quad (8)$$

where the internal energies for the carriers, the LO phonons and the acoustic phonons are given by,

$$U_C = N_C E_g + C_C T_C$$

$$U_O = C_O T_O$$

$$U_A = C_A T_A \quad (9)$$

The relevant physical parameters for the 3TM, their nomenclature and their dependence on temperature and carrier density are given in Table S2.

**Table S2. The parameters of GaAs used in the three-temperature model.**

| Physical Property | Value | Reference |
|---|---|---|
| Heat capacity of carriers  $C_C$ (J/m³ K) | $C_C = 3N_C k_B$ | (21) |
| Heat capacity of LO phonons  $C_{LO}$ (J/m³ K) | $C_{LO} = 3.06 \times 10^5 - 2.4 \times 10^4 (\theta_{LO}/T_O)^{1.94}$  where $\theta_{LO}$ = 344 K | (20, 22) |
| Heat capacity of acoustic phonons  $C_A$ (J/m³ K) | $C_A = 9.17 \times 10^5 - 4.40 \times 10^4 (\theta_D/T_A)^{1.948}$  where Debye temperature $\theta_D$ = 344 K | (21, 22) |
| Thermal conductivity of carriers  $k_C$ (W/m K) | $k_C = 2.5 \times N_C\, k_B{}^2 \tau_m T_C / m^*$  where $m^*$ = 0.066 $m_e$, $\tau_m$ = 0.3 ps | (20, 23) |
| Thermal conductivity of acoustic phonons  $k_A$ (W/m K) | $k_A = 4.0 \times 10^4 / T_A^{1.2}$ | (20, 24) |
| Energy relaxation time between carriers and LO phonons $\tau_{C-O}$ | $\tau_{C-O} = 0.1 \times 10^{-12}$ (s) | (25) |
| Energy relaxation time between carriers and acoustic phonons  $\tau_{C-A}$ | $\tau_{C-A} = 0.5 \times 10^{-12} [1 + (N_C/N_O)^2]$  where $N_O = 2 \times 10^{27}$ (1/m³) | (20, 21) |
| Energy relaxation time between LO and acoustic phonons $\tau_{O-A}$ | $\tau_{O-A} = 8 \times 10^{-12}$ (s) | (20, 26) |
| Absorption coefficient $\alpha_1$ (1/m) | $\alpha_1 = 3.48 \times 10^6 \exp[1.71 \times (x - 1.83)]$<br><br>$x = \hbar\omega + E_g(300) - E_g(T_p)$ (eV) | (24) |
| Auger coefficient $\gamma_A$ (m$^6$/s) | $\gamma_A = 1.0 \times 10^{-43}$ | (27, 28) |
| Band gap $E_g$ (eV) | $E_g(T_p) = 1.575 - 0.15(T_p/300)$ | (24, 29) |
| Linear thermal expansion coefficient $\beta$ | $\beta = 5.7 \times 10^{-6}$ (K$^{-1}$) | (30) |
| Ga-As interatomic distance $\ell$ | $\ell = 4.00$ Å | (31) |
| Sound velocity $v_s$ | $v_s = 4730$ m/s | (32) |

Quantitative analysis for the eutectic growth of alloy phases

1. Laser fluence at 5.5 mJ/cm$^2$

From our studies we could extract some important thermal properties which have not been tabulated in the literature, such as latent heat and heat capacity for the alloy phases during the eutectic growth upon laser heating of the Au/GaAs NWs. To facilitate quantitative analysis of the experimental data including the length reduction of the GaAs NW and the size increment of the bead after each laser pulse, we have employed TTM for gold and 3TM for GaAs to estimate the heat energy absorbed by the NW and the bead upon each laser pulse excitation at a given fluence. According to our simulation, after the 1$^{st}$ laser pulse with a fluence of 5.5 mJ/cm$^2$, the calculated lattice temperatures for the Au bead (566

K) and the GaAs NW (574 K) are above the temperature of ~555 K needed for the $Au_7Ga_2$ phase formation. Therefore, at this fluence those Ga atoms from the disappearing NW volume, corresponding to a length reduction of 25 nm and a hexagonal cross section with a side length of 54 nm, are expected to undergo the reaction with the encapped Au atoms at the tip to form $Au_7Ga_2$ alloy. The excessive heat energy above the temperature (555 K) from the tip and the NW allows us to estimate the latent heat for the alloy formation. Based on the parameters including the molar volumes (Au: $1.0\times10^{-5}$ $m^3$/mol, GaAs: $2.7\times10^{-5}$ $m^3$/mol), the heat capacities (Au: $2.49\times10^6$ $J/m^3$.K, GaAs: $1.76\times10^6$ $J/m^3$.K) (31, 33), the densities (Au: $1.93\times10^4$ $kg/m^3$, GaAs: $5.32\times10^3$ $kg/m^3$, $Au_7Ga_2$: $1.67\times10^4$ $kg/m^3$) (34) and the molar ratio of 7:2 for $Au_7Ga_2$, we obtained the latent heat of 8 kJ/mol for the $Au_7Ga_2$ alloy.

After the 2$^{nd}$ laser pulse at the same fluence but with a larger bead of a mixture of gold and alloy, we observed an additional 3 nm length reduction for the NW. According to the 3TM modeling, we obtained a lattice temperature of 574 K for the GaAs NW. For the top bead, the calculated lattice temperatures were 510 and 567 K when 100% gold and 100% GaAs were assumed in the bead, respectively. Using these temperature values and the molar percentage in the bead, we obtained the absorbed heat from the additional bead volume and from the NW with the values of $1.5\times10^{-14}$ J and $1.1\times10^{-14}$ J, respectively. We defined the net heat as the energy needed to raise the bead for the volume reduction in the NW (3 nm times the hexagonal cross-section area of the NW) from room temperature to 555 K. This net heat was calculated from the overall absorbed heat of the bead and the NW subtracting two parts: first, the heat for raising the temperature to 555 K for the gold molar volume (a 7/2 ratio times the Ga molar volume from the reduction of the NW volume); second, the latent heat of the corresponding Ga molar volume from the reduction of the NW volume. Knowing the density of $Au_7Ga_2$, the specific heat of the $Au_7Ga_2$ alloy was estimated to be 62 J/mol.K.

After the 3rd laser pulse at the same fluence, we observed 1 nm length reduction for the NW. Following the similar approach, a lattice temperature of 574 K for the GaAs NW was obtained. The calculated lattice temperatures of the top bead were 507 and 567 K when 100% gold and 100% GaAs were assumed, respectively. Using the similar procedures outlined above, we extracted the specific heat for the $Au_7Ga_2$ alloy to be 60 J/mol.K, which is close to the value obtained from the 2nd pulse laser heating.

2. Laser fluence at 7 mJ/cm$^2$

After three laser pulses at 5.5 mJ/cm$^2$, the NW length was shortened by 25, 3 and 1 nm sequentially, and the bead stopped to grow. To continue the reaction dynamics, the laser fluence was increased to 7 mJ/cm$^2$. Using the similar simulation approach and the experimental observation of a further 10 nm length reduction of the NW, we obtained the lattice temperatures of 658 K for the GaAs NW, 587 K for a 100% gold bead and 649 K for a 100% GaAs bead, and the overall effective temperature was above the temperature of ~620 K needed for AuGa phase. The absorbed energies from the volume reduction of the NW and from the bead with a mixed composition are $4.8\times10^{-14}$ J and $5.3\times10^{-14}$ J, respectively. To calculate the latent heat of AuGa, one needs to deduct the latent heat of $Au_7Ga_2$ phase and also the energy needed to raise the alloy from room temperature to 620 K for AuGa. These two parts of heat to be subtracted from the overall absorbed laser heat is $2.3\times10^{-14}$ J. Using the density of $1.28\times10^4$ kg/m$^3$ for AuGa (34), its latent heat was estimated to be about 21 kJ/mol.

After the 2nd laser pulse at the same fluence, a 3 nm length reduction of the NW was further observed. Based on the 3TM modeling, the lattice temperature of the GaAs NW is found to be 658 K. For the top bead, the calculated lattice temperatures are 571 K for a 100% Au bead and 645 K for a 100% GaAs bead, respectively. Using these temperature values and the molar percentage in the bead, we estimated the absorbed heat from the additional bead volume and from the NW to be $6.6\times10^{-15}$ J and 4.8×10$^{-15}$ J, respectively. According to the similar procedures mentioned above, we obtained the net heat of 4.65×10$^{-15}$ J. Knowing the density of AuGa, its specific heat of about 41 J/mol.K was obtained.